\documentclass[aps,showpacs]{revtex4}

\usepackage{epsfig}
\usepackage{epsf}
\usepackage{bm}                 % for bold math symbols
\usepackage{amsmath}
\usepackage{amssymb}
\usepackage{dcolumn}
\usepackage{graphicx}

\input amssym.def
\input amssym

\font\bbold=msbm10

\newcommand{\BR}{\mbox{\bbold R}}

\newcommand{\M}{\mathfrak{M}}
\newcommand{\fL}{\mathfrak{L}}
\newcommand{\fS}{\mathfrak{S}}
\newcommand{\NL}{\mathfrak{NL}}
\newcommand{\IN}{{\cal H}_{\rm in}}
\newcommand{\OUT}{{\cal H}_{\rm out}}
\begin{document}

\title{Signaling, Entanglement, and
Quantum Evolution Beyond Cauchy Horizons}

\author{Ulvi Yurtsever} \email{Ulvi.Yurtsever@jpl.nasa.gov}

\affiliation{Quantum Computing Technologies Group, Jet Propulsion 
Laboratory, California Institute of Technology \\ Mail Stop 126-347, 
4800 Oak Grove Drive, Pasadena, California 91109-8099}

\author{George Hockney} \email{George.Hockney@jpl.nasa.gov}

\affiliation{Quantum Computing Technologies Group, Jet Propulsion 
Laboratory, California Institute of Technology \\ Mail Stop 126-347, 
4800 Oak Grove Drive, Pasadena, California 91109-8099}

\date{\today}

\begin{abstract}
Consider a bipartite entangled system half of which falls through the
event horizon of an evaporating black hole, while the other half remains
coherently accessible to experiments in the exterior region. Beyond
complete evaporation, the evolution of the quantum state
past the Cauchy horizon cannot remain unitary, raising the questions:
How can this evolution be described as a quantum map, and how is
causality preserved? What are the possible effects of
such nonstandard quantum evolution maps on the behavior
of the entangled laboratory partner?
More generally, the laws of quantum evolution
under extreme conditions in remote regions (not just in
evaporating black-hole interiors, but possibly near other
naked singularities and regions of extreme spacetime
structure) remain untested by observation, and might conceivably
be non-unitary or even nonlinear, raising the same questions
about the evolution of entangled states.
The answers to these questions are subtle, and are linked in
unexpected ways to the fundamental laws of quantum mechanics. We show
that terrestrial experiments can be designed to probe and constrain
exactly how the laws of quantum evolution might be altered, either by
black-hole evaporation, or by other extreme processes in remote regions
possibly governed by unknown physics.
\end{abstract}

\pacs{03.67.-a, 03.65.Ud, 04.70.Dy, 04.62.+v}

\maketitle

\begin{center}
{\bf \noindent 1. Overview}
\end{center}

Standard proofs
that non-local Bell correlations~\cite{bellcorr} between
parts of an entangled system cannot be used to acausally
signal (transfer information) rely on quantum evolution being
everywhere unitary. However, as Hawking~\cite{hawkingnonunit} first
pointed out when he gave examples of non-unitary but causal
maps for evaporating black holes,
unitarity, a sufficient but not a necessary condition
for causality, may break down in the late stages
of black-hole evaporation. In this paper we ask: When entangled systems
partly cross the event horizons of evaporating black holes
(or Cauchy horizons of other, more general
naked singularities) and partly remain
coherently accessible to experiments outside, what constraints
on their non-unitary, and possibly nonlinear quantum evolution
would ensure causality? and: Can signaling (acausal) evolution be detected
at large distances if it indeed
does take place under the extreme conditions near
naked singularities and evaporating black-hole interiors?
%While conserving probability beyond the domain of dependence
%is a sufficient constraint for an
%un-entangled system, identifying the constraints
%on an entangled system's trans-Cauchy-horizon evolution that
%would guarantee no-signaling turns out to be more subtle.

It turns out, as we will show below,
that linearity (along with probability conservation
and locality) is sufficient to preserve causality;
acausal signaling is possible only with nonlinear maps.
Nonlinear generalizations of quantum mechanics and their
implications for measurement theory and causality have been discussed
by many authors~\cite{nlrefs}; it is not our goal in this paper to
contribute to these formal developments. We adopt the conservative position
that at most a minimal generalization of quantum theory---namely one that
allows for the possibility of
nonlinear quantum maps while keeping the rest of the formalism
intact---is necessary
to understand the implications of
non-standard quantum dynamics for entangled states.
There is, of course, no experimental
evidence for quantum nonlinearity
under local laboratory conditions~\cite{weinberg};
however, whether linearity
continues to hold under extreme conditions such as those
inside evaporating black-holes is a question
yet to be decided by experiment. We will discuss a simple terrestrial
experiment which can probe this question.

Our goal is to show two things: (i) Although any nonlinearity in quantum
mechanics under ordinary laboratory conditions is essentially ruled-out
by local experiments, late stages of black-hole evaporation, and, more
generally, naked singularities, are environments where conditions are
extreme enough (and the local physics is sufficiently uncertain) to
raise the possibility of quantum nonlinearity. And (ii) one can probe
quantum physics in these remote extreme regions of spacetime by {\em
local} terrestrial experiments, such as by the experiment
we propose in detail below. The main argument of this paper
consists of explaining why (a) the proposed experiment is a novel
test of certain generic violations of the linearity of
quantum mechanics {\em at large distances}, (b) that such violations
at-a-distance, though unlikely, are nevertheless {\em not} ruled out by
existing local experiments, and (c) that therefore the proposed
experiment is a compelling test,
since the laws of quantum evolution under extreme
conditions in remote regions (such as in black-hole interiors, near
naked singularities, and other regions where the laws remain untested by
observation, for example, beyond our cosmological event horizon)
{\em might} be nonlinear. 

The experiment we outline in detail in this paper does
not require extremely high energies, sensitivity,
or other ultra-improved technologies to perform:
The ``proof of concept" for our proposed experiment has already been
successfully demonstrated by Mandel et.\ al.~\cite{mandeletal}
more than a decade ago. Our
proposed modification of the Mandel experiment can be performed today,
with a small investment of research effort, by any of the dozens of
quantum optics laboratories around the world.

Throughout the paper, our focus will be on complete
black-hole evaporation as the most likely possible source
of a signal, and the most plausible target, for our experiment.
In fact, in an attempt to restore the unitarity of the evaporation process,
Horowitz and Maldacena~\cite{HM} recently proposed a
boundary-condition constraint for the final quantum state of an
evaporating black hole at its singularity. Gottesman and
Preskill~\cite{GP} have argued that the proposed constraint must lead to
nonlinear evolution of the initial (collapsing) quantum state. Here we
will show that this evolution is a signaling quantum map,
detectable outside the event horizon with the entangled
probe we propose, and making the
Horowitz-Maldacena proposal subject to terrestrial tests.
Independently of this specific example,
our ultimate goal is to convince the reader
that the trans-horizon Bell-correlation
experiment we are proposing is worth doing.
The black-hole emphasis of our argument is grounded in the
view that black-hole
evaporation is the most likely known candidate for new physics
that involves a breakdown in the linearity of quantum mechanics.
Naturally, one could also argue for ``unknown physics" elsewhere
in our future lightcone as a potential target for the experiment;
we foresee, however,
that many readers may find this argument less persuasive.

\begin{center}
{\bf \noindent 2. Black-hole evaporation and non-standard quantum
mechanics}
\end{center}

Why expect the experimentally well-established law of unitary
evolution to break down during black-hole evaporation?
Consider, for definiteness, a pure quantum-field state
which gravitationally collapses to form an evaporating
Schwarzschild black hole (Fig.\,1).
Initially given by $| \Psi_0 \rangle$ on the
(partial) Cauchy surface $\Sigma_0$
in Fig.\,1, the state evolves unitarily
(at least in semiclassical gravity)
during and after gravitational collapse:
at any intermediate time slice
\begin{figure}[htp]
%\vspace*{-1in}
\hspace{1.6in}
\centerline{
\input epsf
\setlength{\epsfxsize}{5.100in}%{3.400in}
\setlength{\epsfysize}{4.461in}%{2.974in}
\epsffile{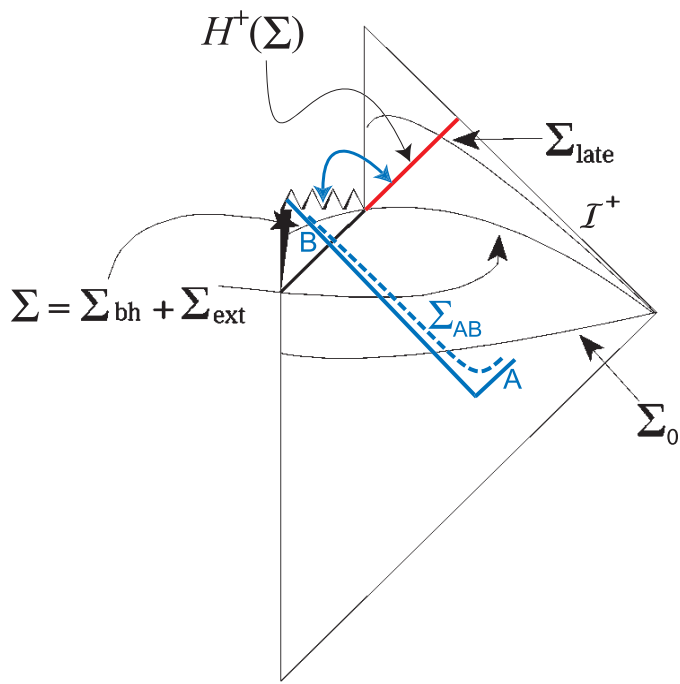}
}
\vspace{-0.90in}
\caption[figure]
{\label{fig:figure1}
Conformal diagram illustrating
the causal structure of a spacetime with an evaporating black hole
(vertical lines on the left depict the axes of rotational symmetry).
[The causal geometry illustrated by the superimposed blue drawing
refers to the paragraph immediately preceding Sect.\,4 below.]
The spacelike hypersurface $\Sigma_0$
passes through the collapsing star before the black
hole has formed, $\Sigma$ is a surface through the black hole
just before it evaporates, and $\Sigma_{\rm late}$
is a surface at late times, after
complete evaporation. The red line illustrates the Cauchy horizon
$H^{+}(\Sigma )$ for $\Sigma$ or $\Sigma_0$;
it is the future null cone of the ``point" (really a singularity)
of complete evaporation.
Because evaporation is largely thermal, quantum evolution
through $H^{+}(\Sigma )$ from the time slice
$\Sigma_0$ to the slice $\Sigma_{\rm late}$
cannot be described as a unitary map.}
\vspace{-0.10in}
\end{figure}
$\Sigma$, it can be written as $
|\Psi_{\Sigma}\rangle = U_{\Sigma \, \Sigma_0 }
|\Psi_0 \rangle$, where $U_{\Sigma \, \Sigma_0 }$
is the unitary time evolution operator acting on the Fock space
of field states. An external observer
in the asymptotically flat region outside the event horizon has no
causal communication with the interior $\Sigma_{\rm bh}$; she
would describe the state of the quantum field by the reduced density matrix
\begin{equation}
\rho_{\rm ext} = {\rm Tr}_{\Sigma_{\rm bh}} \, |\Psi_\Sigma \rangle
\, \langle \Psi_\Sigma | \; 
\end{equation}
obtained by tracing over the interior field degrees
of freedom inside the horizon.
As the black hole settles down to a stationary
state on the time slice $\Sigma$,
the mixed state $\rho_{\rm ext}$ can be shown (via non-trivial
calculation~\cite{bhentropy}) to approach precisely a
thermal state $\rho_H$ at the Hawking temperature $T_H = \hbar c^3
/(8 \pi k_B G M)$, where $M$ is the hole's mass.
As long as the back action of
the Hawking radiation on spacetime is negligible (an eternal black hole),
matter remains in the pure state $|\Psi_{\Sigma} \rangle$,
which unitarily evolves to become entangled with its collapsed half
inside the emerging event horizon. But what happens at late times,
after this back action eventually destroys the black hole completely?
In semiclassical gravity, it is impossible
to escape the conclusion that the state $\rho_{\rm late}$ of the field
on the late time slice $\Sigma_{\rm late}$ (Fig.\,1) is mixed:
$\rho_{\rm late} \approx \rho_H$. The resulting
evolution $|\Psi_0 \rangle
\longmapsto \rho_{\rm late}$ cannot be unitary, as it maps pure states
into mixed states.
This inevitable breakdown of unitarity
can only be avoided by postulating a remnant that
persists at late times, continuing to carry the correlations ``lost"
in the state $\rho_{\rm late}$ by remaining entangled with the outgoing
Hawking radiation.

The lesson we draw is: compared to the conditions encountered in local
laboratory physics, conditions in the interiors
of evaporating black holes are so extreme that
the ordinary laws of quantum evolution
may be profoundly altered~\cite{ftnote1}.
What kinds of non-unitary quantum dynamics might govern entangled
multi-partite systems as their subsystems
cross the Cauchy horizons of evaporating black holes?
We argue that this dynamics must be probability-preserving,
it can be (generally) nonlinear, and it must be local.
The class of non-unitary maps
(``superscattering operators")
discussed by Hawking~\cite{hawkingnonunit}
is obtained via the additional constraint of linearity.
We will show that linearity (along with probability conservation
and locality) is sufficient to preserve causality~\cite{ftnote2};
acausal signaling is possible only with nonlinear maps.
There is, of course, no experimental
evidence for quantum nonlinearity
under local laboratory conditions~\cite{weinberg};
however, whether linearity
continues to hold under the extreme conditions
of evaporating black-hole interiors is a question
yet to be decided by experiment. Remarkably, a simple terrestrial
experiment can be designed to probe this question as we now discuss.

\begin{center}
{\bf \noindent 3. The trans-horizon Bell-correlation experiment}
\end{center}

Consider the optical setup schematically illustrated in Fig.\,2,
a straightforward modification of a well-known Bell-correlation
experiment by Mandel et.\ al.~\cite{mandeletal}. The pump beam
(typically from the output of a uv-argon laser) is split
into two beams which
interact with two separate nonlinear crystals to produce
correlated photons in two pairs of idler and signal beams, labeled
$u$, $x$, and $d$, $e$, respectively. The key feature in the
design of the experiment is the alignment of the first signal beam
$x$ with the second signal beam $e$, which makes photon number-states
in the beams (modes) $x$ and $e$ indistinguishable (in practice,
the alignment needs to be accurate only to within the transverse
laser coherence length). In the actual experiment
the first signal beam $x$ may pass through the second
nonlinear crystal as a consequence of its alignment with $e$,
but its probability of further down-conversion,
proportional to $|V f_1 f_2 |^2$, is negligible since $|f_i |\ll 1$
and $|V f_i |\ll 1$, $i=1, \; 2$, where $V$ is the dimensionless
amplitude of each of the two pump-beam pulses
(photon number $\propto |V|^2$).
We shall assume that both nonlinear crystals produce down-converted
photons in a fixed (linear) polarization state.
The quantum state output by this
configuration belongs to the Hilbert space ${\cal H} \equiv {\cal H}_u
\otimes {\cal H}_d \otimes {\cal H}_e$, where the ``up" and
``down" idler-beam Hilbert spaces are generated by the
orthonormal basis states
\begin{equation}
{\cal H}_u \equiv < \{ | 0 \rangle_u , \; |1 \rangle_u \} > \; ,
\;\;\;\; {\cal H}_d \equiv < \{ | 0 \rangle_d , \; |1 \rangle_d \} >
\; ,
\end{equation}
and the ``escaping" signal-beam Hilbert space is generated by the
basis states
\begin{equation}
{\cal H}_e \equiv < \{ | 0 \rangle_e , \; |1 \rangle_e
 , \; |-1 \rangle_e  \} > \; ,
\end{equation}
where $|0\rangle$ denotes the vacuum, $|1 \rangle$ denotes the
single-photon state in the original
(linear) polarization mode produced by the
down-conversion, and $|-1\rangle$ denotes the single-photon state
in the orthogonal polarization mode,
which is mixed into ${\cal H}_e$ by the polarization
rotator (with complex coefficients $a$,
$b$) placed along the signal beam $x$ (Fig.\,2). The output state can
be written as
\begin{eqnarray}
|\psi\rangle & = &
[\; |0\rangle_u |0\rangle_d |0\rangle_e \nonumber \\
& + &
V f_1 e^{i \phi} \, |1 \rangle_u |0\rangle_d \, (\; a \, |1 \rangle_e
\, + \, b \, |-1\rangle_e \; ) \nonumber \\
& + & V f_2 \, |0\rangle_u |1\rangle_d |1\rangle_e \; ]\, / \, N \; \; ,
\end{eqnarray}
where $|a|^2 + |b|^2 = 1$, and $N$ is the normalization factor
\begin{equation}
N \equiv \sqrt{1 + |V|^2 (|f_1 |^2 + |f_2 |^2 )} \; .
\end{equation}
Notice that the contributions from the signal beam $x$ and from the
signal beam $e$ are coherently superposed in the output state
$|\psi\rangle$ along the ${\cal H}_e$-direction in ${\cal H}$;
this is the key consequence of aligning the two
signal beams.
\begin{figure}[htp]
%\vspace*{0.3in}
\hspace{3.0in}
\centerline{
\input epsf
\setlength{\epsfxsize}{5.46in}%4.200
\setlength{\epsfysize}{4.2185in}%3.245
\epsffile{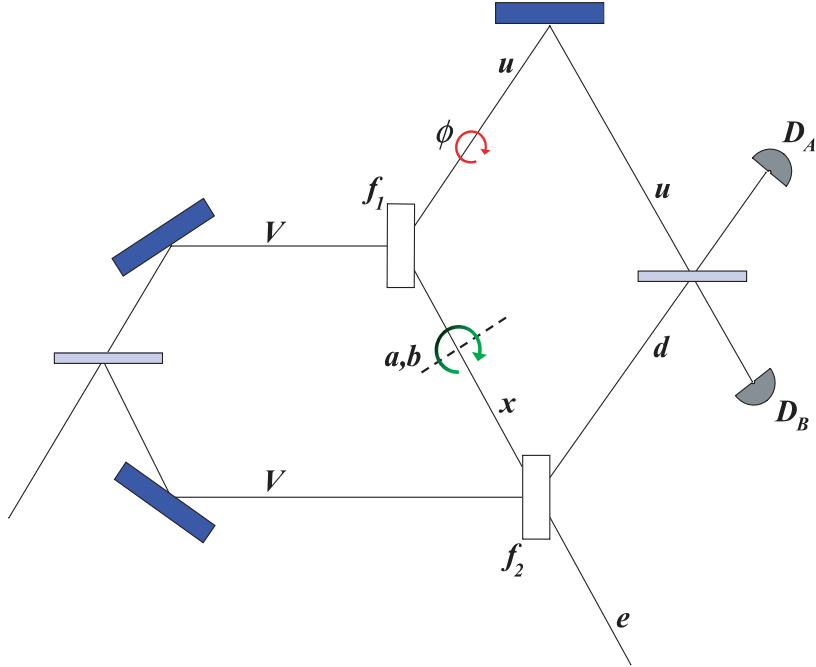}
}
\vspace{-0.60in}
\caption[figure]{\label{fig:figure2}
The Zou-Wang-Mandel interferometer~\cite{mandeletal} for the trans-horizon
Bell-correlation experiment. Gray rectangles are
50:50 beam splitters, white rectangles are the two nonlinear parametric
down-converting crystals with efficiencies $f_1$, $f_2$; blue
rectangles are mirrors. A phase delay is placed on the idler
beam labeled $u$, and an adjustable-angle
polarization rotator is placed on the signal beam $x$ which is aligned
with the second signal beam labeled $e$. Since both pump beams
are blocked past the nonlinear crystals, the output state lies
in the Hilbert
space ${\cal H}_{u} \otimes {\cal H}_{d} \otimes {\cal H}_{e}$ of
the signal and idler beams;
it is monitored by the single-photon detectors $D_A$ and $D_B$.}
\vspace{-0.00in}
\end{figure}

The experiment consists of monitoring the entangled output
$|\psi\rangle$ at the two single-photon detectors $D_A$ and $D_B$.
For the purposes of our essentially conceptual discussion in this
paper, experimental inaccuracies and noise
(detector inefficiencies, dark-count rates, $\ldots$)
are not relevant, and we will defer their discussion to
a forthcoming paper~\cite{moretocome}. Thus, measurement by the perfect
detector $D_A$ is equivalent to the projection $P_A = P_{\alpha}
\otimes \mathbb{I}_e$, where $\alpha \in {\cal H}_u \otimes {\cal H}_d$
is the vector
$\alpha = (|0\rangle_u |1 \rangle_d
+ i |1\rangle_u |0 \rangle_d )/{\sqrt{2}}$,
and a measurement (click) at $D_B$ is equivalent to the projection
$P_B =  P_{\beta} \otimes \mathbb{I}_e$,
where $\beta = (|1\rangle_u |0 \rangle_d
+ i |0\rangle_u |1 \rangle_d )/{\sqrt{2}}$.
Calculation [using $p_{A,\; B}  =  {\rm Tr}
( P_{A,\; B}  |\psi\rangle\langle \psi | )
 = \| P_{A, \; B} |\psi \rangle \|^2$] shows that
the probabilities $p_A$ and $p_B$ of clicks at detectors $D_A$ and
$D_B$, respectively, are given by
%\begin{eqnarray}
%p_A & = & {\rm Tr}
%\, (\, P_A \, |\psi\rangle\langle \psi | \, )
%\, = \, \| P_A |\psi \rangle \|^2 \; \; , \nonumber \\
%p_B & = & {\rm Tr}
%\, (\, P_B \, |\psi\rangle\langle \psi | \, )
%\, = \, \| P_B |\psi \rangle \|^2\; \; .
%\end{eqnarray}
%Calculation using Eqs.\,(4)--(8) gives
\begin{eqnarray}
p_A & = & \frac{|V|^2}{2 \, N^2} \left[ \,
|f_1 |^2 + |f_2 |^2 \, + \, 2 \, \Re \, ( \,
i \overline{f_1 }f_2 \bar{a} e^{-i \phi }
\, ) \, \right] \; , \nonumber \\
p_B & = & \frac{|V|^2}{2 \, N^2} \left[ \,
|f_1 |^2 + |f_2 |^2 \, - \, 2 \, \Re \, ( \,
i \overline{f_1 }f_2 \bar{a} e^{-i \phi }
\, ) \, \right]  \; .
\end{eqnarray}
The important feature in Eqs.\,(6) is the interference term in brackets
following the real-part sign $\Re$. Notice that the interference is
oscillatory in the controlled phase delay $\phi$ and
depends sensitively on the polarization angles $(a,b)$.

But how does the interference depend on the evolution of
the probe beam $e$ which escapes to infinity? Let
$\rho_{ud} \equiv {\rm Tr}_e |\psi\rangle \langle \psi |$
be the output state projected on the ``laboratory"
Hilbert space ${\cal H}_u \otimes {\cal H}_d$.
It is straightforward to show that
the detection probabilities $p_{A,\; B}$ can be
alternatively computed via the expressions $p_{A, \; B}
= {\rm Tr}_{ud} ( P_{\alpha , \; \beta} \, \rho_{ud})$.
This result is, of course,
valid much more generally: the expectation value of
any observable
$O = O_{ud} \otimes \mathbb{I}_e$ (i.e.\ one local to the
Hilbert space ${\cal H}_u \otimes {\cal H}_d$) depends only on the
reduced state projected on ${\cal H}_u \otimes {\cal H}_d\;$:
\begin{equation}
{\rm Tr}\, [ \, ( O_{ud} \otimes \mathbb{I}_e )\,
|\psi\rangle\langle\psi |\, ] = {\rm Tr}_{ud}\,
[\, O_{ud}\, {\rm Tr}_e |\psi\rangle\langle\psi | \, ] \; .
\end{equation}
Now suppose that the output state $|\psi\rangle$ undergoes a local
quantum evolution (local in the sense that
${\cal E} = {\cal E}_{ud} \otimes {\cal E}_e$)
\begin{equation}
{\cal E} = {\cal E}_{ud} \otimes {\cal E}_e \; : \;
|\psi \rangle
\langle \psi | \mapsto
({\pmb 1}_{ud} \otimes {\cal E}_e )(|\psi \rangle
\langle \psi |) \; ,
\end{equation}
where ${\cal E}_e$ is an arbitrary, completely-positive,
linear quantum map on ${\cal H}_e$-states which
is probability preserving, with Kraus representation:
\begin{equation}
{\cal E}_e  : \rho \mapsto
\sum_{j} E_j \rho {E_j}^\dagger \; , \;\;\;\;\;
\sum_j {E_j}^\dagger E_j = \mathbb{I}_e \; ,
\end{equation}
where $E_j$ are otherwise arbitrary linear operators on ${\cal H}_e$.
For any state $\rho$ on ${\cal H}$ [including the output state
$\rho = |\psi\rangle\langle\psi | $ of Eq.\,(4)], by expanding $\rho$
in the form $\rho = \sum_\mu c_\mu \,
{\rho_{ud}}^{(\mu )} \otimes {\sigma_e}^{(\mu )}$,
$c_\mu \in \mathbb{R}$, it is straightforward
to prove the identity
\begin{equation}
{\rm Tr}_e \, [ \, ({\pmb 1}_{ud} \otimes {\cal E}_e ) \rho \, ]
= {\rm Tr}_e \rho
\end{equation}
for any {\em linear} map ${\cal E}_e$ of the form Eq.\,(9).
In view of Eq.\,(7), Eq.\,(10) is the
expression of causality (no-signaling;
compare Eq.\,(15) below): As long as the evolution of the
probe beam $e$ remains linear and probability-conserving, the
interference pattern of the laboratory beams does not depend on what
happens to $e$. The detection probabilities $p_A$ and $p_B$ are given by
Eqs.\,(6) whether $e$ evolves unitarily, is absorbed in a beam block, or
otherwise gets entangled with the rest of the universe.

By contrast, suppose
that the beam $e$ undergoes a {\em nonlinear},
probability-conserving evolution. As an example, consider
the evolution proposed
in~\cite{HM} for evaporating black holes, whose
action on any state $\rho \in {\cal H}$ is given by
(see Sect.\,5 below for a detailed discussion of this map class)
\begin{equation}
{\cal E} \; : \; \rho \longmapsto
\frac{{\pmb 1}_{ud} \otimes {\cal T}_e \; ( \rho)}
{{\rm Tr} [{\pmb 1}_{ud} \otimes {\cal T}_e \; ( \rho)]} \; ,
\end{equation}
where ${\cal T}_e$ denotes the linear transformation (not a quantum map)
${\cal T}_e : \rho_{e} \mapsto T_e  \rho_{e}  {T_e}^{\dagger}$
on states $\rho_e$ of ${\cal H}_e$, and
$T_e : {\cal H}_e \rightarrow {\cal H}_e$
is an arbitrary nonsingular linear transformation.
%\begin{equation}
%\kappa |0\rangle_e + \delta |1\rangle_e
%+ \gamma |-1\rangle_e \longmapsto 
%\kappa \, |0\rangle_e + \sqrt{|\delta |^2 + |\gamma |^2 }
%\, |1\rangle_e \; 
%\end{equation}
%on ${\cal H}_e$~\cite{ftnote3}. This nonlinear evolution maps the
%output state $|\psi\rangle$ of Eq.\,(4) into the state $|\psi_{\rm
%nl}\rangle$ given by
%\begin{eqnarray}
%|\psi\rangle \mapsto |\psi_{\rm nl}\rangle
%& = & \left[ \, |0\rangle_u |0\rangle_d |0 \rangle_e
%\, + \, |V f_1 | \; |1\rangle_u |0 \rangle_d |1\rangle_e \right.
%\nonumber \\
%& + & \left. |V f_2 | \; |0\rangle_u
%|1 \rangle_d |1\rangle_e \, \right] \, / \, N \; \; .
%\end{eqnarray}
%Calculation of the new detector-click probabilities ${p^\ast}_{A,B}
%\equiv {\rm Tr} (P_{A,B} \,
%|\psi_{\rm nl}\rangle\langle\psi_{\rm nl} | )$ gives
%\begin{equation}
%{p^\ast}_{A,B} = \frac{|V|^2}{2 \, N^2} \left[ \,
%|f_1 |^2 \, + \, |f_2 |^2 \, \right]  \; .
%\end{equation}
For simplicity, let us choose $T_e$ in the form
\begin{equation}
T_e = 
  \begin{pmatrix}
   1 & 0 & 0 \cr
   0 & 1 & 0 \cr
   0 & -1 & 1
  \end{pmatrix}
\end{equation}
in the $\{|0\rangle_e , \; |1\rangle_e , \; |-1\rangle_e \}$ basis
of ${\cal H}_e$.
After the incoming state
$\rho=|\psi\rangle \langle \psi|$ is transformed
into ${\rho}^{\ast} \equiv {\cal E}(\rho )$
according to the nonlinear evolution ${\cal E}$
given by Eqs.\,(11)--(12), the probabilities of
detection at the local detectors can be re-calculated using the
equations ${p^\ast}_{A, \; B}
= {\rm Tr} ( P_{A , \; B} \, {\rho^{\ast}})$.
The result for the interference signal $p_A - p_B$ is:
\begin{equation}
{p^\ast}_A - {p^\ast}_B = \frac{2 \, |V|^2}{{N^\ast}^2}\;
\Re \left[ \,
i \overline{f_1 }f_2 \, (\bar{a} - \bar{b}) \, e^{-i \phi }
\, \right] \; ,
\end{equation}
where ${N^\ast}^2 = 1 + |V|^2 [\, |f_2 |^2 + |f_1 |^2 (|a-b|^2+|b|^2
)\, ]$ is the new normalization factor.
Observing a signal like Eq.\,(13) would represent a clean
detection of the nonlinear map $\cal E$ [Eqs.\,(11)--(12)]
by our interferometer, since, e.g., the new null and maximum
of the interference with respect to the polarization-rotator
angle $\theta \equiv \arctan |b/a|$ are both shifted by $45^{\circ}$
compared to Eqs.\,(6).
In general, the interference signal $p_A - p_B$ as a function of $\phi$ and
$\theta$, the fundamental observable in our proposed experiment,
constitutes a rich 2D
data set sensitive to almost any nonlinear
evolution map affecting the probe beam $e$.

It is important to note that quantum evolution in the presence of signaling 
maps is incompatible with standard
Cauchy evolution (in globally hyperbolic spacetimes, or within the domain
of dependence of a partial Cauchy surface when global hyperbolicity
fails). In fact, the phenomenon of signaling
is equivalent to the well-known ambiguity that arises when one attempts
to carry out time evolution from one partial Cauchy surface
(i.e.\ a spacelike surface no causal curve intersects
more than once) to another
using the usual existence and uniqueness properties of the
standard Cauchy problem~\cite{Jacobson}.
Instead, one makes consistent predictions in the
presence of signaling based on principles similar to the concept of
``self-consistency" which provides for consistent
predictions in the presence of closed timelike curves~\cite{CTC}.
More specifically, the principle of self consistency implies the following
algorithm for the consistent evolution of distributed
quantum states in the presence of signaling maps:
Suppose $P_B$ is a spacetime point where nonlinear (signaling)
quantum evolution takes place, $AB$ is an entangled system whose
wavefunction is localized around worldlines of $A$ and $B$,
and the worldline of $B$ passes through the point $P_B$. Then,
for self consistency, the effect of the nonlinear evolution at
$P_B$ must extend everywhere in
spacetime outside the past null cone of $P_B$. In predicting the
evolution of the joint quantum state of $AB$, one would use the
standard unitary evolution from one partial Cauchy surface
to the next, and, as soon as
there exists at least one partial Cauchy surface
passing through $A$ and $P_B$ [i.e.\ as soon as $A$
enters the complement of $I^- (P_B)$,
the chronological past of $P_B$], apply the nonlinear evolution
map to the entire state of $AB$. Of course this leads to a ``teleological"
influence of $P_B$ on the entire state, but this
property is the characteristic
feature of signaling maps, and it is also because of this
feature that signaling maps can be used to
transmit information acausally.

To apply the evolution principle discussed above to
the problem corresponding to our trans-horizon
Bell-correlation experiment, let $A\equiv \{u,d\}$ and $B\equiv \{ e \}$
for the output state Eq.\,(4) as the probe beam $e$
is directed into the event horizon of a
black hole. Suppose the evaporation of the hole leads to
a quantum map ${\cal E}_{AB}$ with
a signaling nonlinear component.
Would the nonlinearity cause a detectable
shift in the interference patterns of $u$ and $d$?
As discussed in the above paragraph,
and in accordance with standard quantum field
theory, the evolution of $|\psi\rangle$ is
described by ${\cal E}_{AB}$ when and only when
the subsystems $A$ and $B$ are contained in a partial
Cauchy surface $\Sigma_{AB}$.
The blue diagram in Fig.\,1 depicts such a surface
$\Sigma_{AB}$ for the causal geometry of the proposed experiment.
If the ultimate causal structure of the evaporating
quantum black hole remains the same
as given by the classical metric (Fig.\,1),
the singularity is a final boundary, $e$ will
propagate unitarily before it disappears into the singularity,
and no signal will be produced (effectively
unitary ${\cal E}_{AB}$). If, on the other hand,
$e$ re-emerges as Hawking radiation following evaporation
[i.e.\ if the singularity is effectively a part of $H^{+}(\Sigma
)$ in the quantum spacetime], then a detectable signal will result.
Conversely, the likely null outcome of the experiment
can be used to place precise upper limits on
the strength of any signaling nonlinear component in the
effective quantum map ${\cal E}_{AB}$ of evaporating black
holes (see Sect.\,6 below for more discussion on this and other
feasibility questions for our proposed experiment).

\begin{center}
{\noindent
\bf 4. General theory of nonlinear quantum evolution maps}
\end{center}

If black-hole evaporation compels us to treat linearity
as a property to be tested by experiment rather than as a
universal axiom of quantum mechanics, what
properties must hold for the most general class of
quantum maps governing quantum evolution everywhere?
We now turn briefly to the mathematical description
of this generalized class
of maps. Full details will be found
in the forthcoming~\cite{moretocome}.

A black-hole event horizon divides spacetime naturally
into two distinct regions: the
external, asymptotically-flat region, $A$, and the interior region beyond
the event horizon, $B$, where the causal future of $B$ does not intersect
$A$. The same causal separation is natural for the more general
problem where $B$ represents a distant region with a
naked singularity or other extreme structure
(e.g.\ beyond our cosmological event horizon), and $A$ represents the
local spacetime neighborhood.
Consider now a general quantum evolution map, defined as a map
from the space of all states (density matrices)
of the joint Hilbert
space of $A$ and $B$ (simply $AB$) into the (linear) space of symmetric
operators on the same Hilbert space. There are four key
properties such an evolution map may satisfy:

\begin{itemize}
\item {\em Locality}: The action of the map at
$A$ does not depend on the state of the
system as seen at $B$, and vice versa; i.e., the {\em form}
of the map's action in one region does not depend on what the state of the
system is in the other region. 

\item {\em Probability conservation}:
The evolution sends density matrices of unit
trace (normalized with unit probability) to density matrices of unit
trace. 

\item {\em Causality} (no-signaling condition):
The evolution map cannot be used to
signal between $A$ and $B$ when classical
communication between $A$ and $B$ is not allowed.

\item {\em Linearity}: The evolution map is the restriction to states (i.e. to
density matrices) of a linear transformation on the linear space of all
symmetric operators in the joint Hilbert space of $AB$. (States
are the positive symmetric operators of trace 1; a nonlinear
subset of the space of all symmetric operators.)
\end{itemize}

Note that standard, local unitary quantum maps satisfy
all four properties. We contend that while
the first two conditions, namely locality and probability conservation,
are indispensable for any physical evolution map, the last two
conditions are not. We view locality as indispensable because relaxing it
would amount to assuming communication between the physical ``agents"
implementing the evolution maps at $A$ and $B$,
an assumption which makes acausal signaling possible even with
standard unitary evolution maps, and which contradicts the
presumed causal separation between the regions $A$ and $B$.
Probability conservation is
indispensable because its violation would be easily detectable in almost
every experiment which sends physical signals from the external region
$A$ into the region $B$; since no local experiment has ever detected such
violations, they are ruled out on observational grounds.
As for causality, we take the position that
causality is a higher level notion, which should be
derived as a theorem from the other more primitive laws of quantum mechanics;
when it cannot be so derived, ruling out violations of causality becomes
an experimental question. The same conclusion applies (with
even more force) to linearity.

Given a causally separated bi-partite quantum system $AB$ with
(finite-dimensional) Hilbert space ${\cal H}
\equiv {\cal H}_A \otimes {\cal H}_B$, let $W({\cal H})$
be the real vector space of symmetric operators on $\cal H$,
and $S({\cal H}) \subset W({\cal H})$ 
the set of all states (positive, symmetric operators of unit trace).
We propose that the set of physical quantum maps $\M ({\cal H}) $
should consist of all smooth maps
${\cal E}_{AB} : W({\cal H}) \rightarrow W({\cal H})$
which map $S({\cal H})$ into $S({\cal H})$ (conserve probability)
and which satisfy an appropriate condition for locality.
What form should this locality condition take for
a general nonlinear map? The most general definition of locality
for a {\em linear} bi-partite quantum map ${\cal E}_{AB}$ stipulates
\begin{equation}
{\cal E}_{AB}= {\rm Tr}_{RS} \, \left(
{\cal E}_{AR} \otimes {\cal E}_{BS} \right) \; ,
\end{equation}
where $R$ and $S$ are auxiliary systems co-located with $A$ and $B$,
respectively, the joint system $RS$ is allowed to have
prior-established entanglement (see Eqs.\,(15)--(18) in~\cite{preskill}),
and ${\cal E}_{AR}$, ${\cal E}_{BS}$ are arbitrary linear (positive and
complete) quantum maps acting on the Hilbert spaces ${\cal H}_{AR}$
and ${\cal H}_{BS}$, respectively. The condition Eq.\,(14) is equivalent
to the statement
\begin{equation}
{\cal E}_{AB} = \sum_{\alpha} d_{\alpha} \; {\cal E}_{A \alpha}
\otimes {\cal E}_{B \alpha} \; ,
\end{equation}
where the coefficients $d_{\alpha}$ are real numbers ($\sum_{\alpha}
d_{\alpha} = 1$), and ${\cal E}_{A \alpha}$ and ${\cal E}_{B \alpha}$
are linear quantum evolution maps on ${\cal H}_{A}$ and ${\cal H}_{B}$,
respectively. For nonlinear maps, however, the tensor product
operation is not well defined, and has no canonical generalization.
Consequently, for a general (possibly nonlinear) map
we adopt the more stringent (restrictive) {\em locality condition}
\begin{equation}
{\cal E}_{AB} (\rho \otimes \sigma ) = {\cal E}_A
(\rho ) \otimes {\cal E}_B (\sigma ) \; 
\end{equation}
for {\em all} $\rho \in S({\cal H}_A )$
and $\sigma \in S({\cal H}_B )$, where ${\cal E}_A$
and ${\cal E}_B$ are {\em fixed} quantum maps in $\M ({\cal H}_A )$
and $\M ({\cal H}_B )$ (called the {\em local components}
of ${\cal E}_{AB}$) that depend only on ${\cal E}_{AB}$.
When ${\cal E}_{AB}$ is a linear map, the locality condition
Eq.\,(16) implies that ${\cal E}_{AB}$ must be
in the form of a tensor
product of linear maps ${\cal E}_{AB} = {\cal E}_A \otimes {\cal E}_B$,
where the linear maps ${\cal E}_A$ and ${\cal E}_B$
are the local components of ${\cal E}_{AB}$. Clearly, all linear maps
local according to Eq.\,(16) are local according to the
definition Eqs.\,(14)--(15), but not vice versa.

The condition for a map
${\cal E}_{AB} \in \M$ to be signaling (non-causal) is precisely
that for {\em some} $\rho_{AB} \in S({\cal H})$
\begin{equation}
{\rm Tr}_B \, [ \, {\cal E}_{AB} (\rho_{AB}) \, ]
\neq {\cal E}_A \, [ \, {\rm Tr}_B ( \rho_{AB}  ) \, ] \; ,
\end{equation}
where ${\cal E}_A$ is the local $A$-component of ${\cal E}_{AB}$.
It is easy to prove using Eq.\,(16) [or, more generally,
Eq.\,(15)] that all local linear ${\cal E}_{AB}$
are causal [non-signaling; cf.\ Eq.\,(10)]: Consider any decomposition
$\rho_{AB}=\sum_{i} \, c_i \, \rho_{Ai} \otimes \sigma_{Bi}$, where
$\rho_{Ai}$ and $\sigma_{Bi}$ are normalized states. Since
any local linear ${\cal E}_{AB}$ is of the form
${\cal E}_{AB}={\cal E}_A \otimes {\cal E}_B$,
\begin{eqnarray}
{\rm Tr}_B \, [ \, {\cal E}_{AB} (\rho_{AB}) \, ] & = &
\sum_i \, c_i \, {\rm Tr}_B [
{\cal E}_{AB} (\rho_{Ai} \otimes \sigma_{Bi})] \nonumber \\
& = & \sum_i \, c_i \, {\rm Tr}_B [
{\cal E}_{A} (\rho_{Ai}) \otimes {\cal E}_B (\sigma_{Bi}) ]
=  \sum_i \, c_i \,
{\cal E}_{A} (\rho_{Ai}) \nonumber \\
& = & {\cal E}_A \left(
\sum_i \, c_i \, \rho_{Ai} \right)
= {\cal E}_A \, [ \, {\rm Tr}_B ( \rho_{AB}  ) \, ] \; .
\end{eqnarray}

Note also
that locality [Eq.\,(16)] explains why phase-coherent
entanglement of
$AB$ is essential to detect any non-causal influence of ${\cal E}_{AB}$
at $A$ when, for example, $B$ is inside the
event horizon and $A$ is in the
exterior region of a black hole:
Any system (e.g., starlight) entangled with the external world
will give rise to a decohered input state having the product form $\rho_{AB}
= \rho_A \otimes \sigma_B$, and evolution of such product states
cannot satisfy the signaling
condition Eq.\,(17) because of the locality constraint Eq.\,(16).
Experiments must carefully preserve
phase-coherence of entanglement (as proposed in Fig.\,2)
to be able to detect signaling. We will now give a
more detailed explanation of this point.

A significant component of our main argument, namely, a push for
performing the trans-horizon Bell-correlation experiment, is
that nonlinear maps of the kind this experiment is sensitive to
would not have any observable effects that would have been visible in other
experiments. There are, naturally, many other systems, such as stars,
that produce multi-partite entangled quantum states,
where part of the system (in the case of a star, the emitted light)
might travel to distant regions which may possibly involve
nonlinear quantum evolution maps.
In order to make the argument that nonlinear quantum mechanics ``at
large distances" is not already ruled out by existing experiments,
we need to show why such states are not effective probes for the
putative nonlinearities. The crucial property of entangled
states produced by systems such as
stars is that they have completely random relative phases. For
example, assuming, for conceptual simplicity,
two-dimensional Hilbert spaces ${\cal H}_A$, corresponding
to the emitters, and ${\cal H}_B$,
corresponding to the emitted light, the ``starlight states"
are of the form:
\begin{equation}
|\psi \rangle
=\frac{1}{2}
(|00\rangle + |01\rangle e^{i \phi_1}
+ |10\rangle e^{i \phi_2} + |11\rangle e^{i \phi_3} )
\end{equation}
where $|00\rangle$ etc.\ denotes $|0\rangle_A \otimes |0\rangle_B$ etc.
Therefore, effectively, the input state (in ${\cal H}_A \otimes {\cal
H}_B$) is given by the average:
\begin{equation}
\frac{1}{(2 \pi )^3}
\int_0^{2 \pi} \int_0^{2 \pi} \int_0^{2 \pi}
d\phi_1 \, d \phi_2 \, d \phi_3 \; |\psi \rangle \langle \psi |
= \frac{1}{4} \; {\pmb 1}_A \otimes {\pmb 1}_B \; .
\end{equation}
An alternative derivation of the same conclusion Eq.\,(20)
may be given as follows: The systems $A$ (emitters) and $B$ (starlight)
are entangled with a third system, a reservoir representing the
``environment." In the larger Hilbert
space of the full system including the environment,
 starlight states correspond to states of the form
\begin{equation}
|\psi \rangle
=\frac{1}{2}
(|00\rangle |e_0\rangle + |01\rangle |e_1\rangle
+ |10\rangle |e_2\rangle + |11\rangle |e_3 \rangle )
\end{equation}
where the $|e_i\rangle$ denote orthonormal states of the environment.
In other words, the ``probe beam" is decoherent with the environment.
However, it's not just the probe beam which is entangled with the environment,
but the entire state of the ``apparatus," which in this
example corresponds to not only the starlight ($B$),
but also whatever other system (the sources $A$)
in the star that is producing the starlight. (In our proposed
Bell-correlation experiment of Fig.\,2,
the apparatus $AB$ is a three-partite system; so this bi-partite example
is a bit simplified, but the main conclusion remains the same).
Again, when traced over the environment, the input state becomes
\begin{equation}
{\rm Tr}_{\rm Env} \; |\psi\rangle \langle \psi |
= \frac{1}{4} \; {\pmb 1}_A \otimes {\pmb 1}_B \; ;
\end{equation}
which is as before the maximally mixed state (a maximally mixed
state is always in product form).

In the above very simple example, we assumed that there is no preferred
orthonormal basis state in ${\cal H}_A \otimes {\cal
H}_B$ for a quantum state $|\psi\rangle$ produced by a random
process. In general, there may exist ``superselection sectors"
determined by conserved quantities (such as energy or charge), and basis
states in different sectors (e.g.\ at different energy levels) may have
different weights. For example, a thermal state has this property.
For ``thermal" starlight, the system state would have to be,
instead of Eq.\,(21), an input state $| \psi \rangle$
which when traced over the environment takes the form
\begin{equation}
{\rm Tr}_{\rm Env} \; |\psi\rangle \langle \psi |
= \frac{1}{ Z} \sum_{j, \; k} e^{- \beta ({E}_{jA}
+ {E}_{kB} ) }
\; \frac{{\pmb 1}_{jA}}{d_{jA}} \otimes
\frac{{\pmb 1}_{kB}}{d_{kB}} \;
\end{equation}
where $Z \equiv \sum_{j, \, k} e^{- \beta ({E}_{jA}
+ {E}_{kB} ) }$, and ${\pmb 1}_{jA}$,
${\pmb 1}_{kB}$ denote the identity operators in the respective
energy eigenspaces (of dimensions
$d_{jA}$ and $d_{kB}$, and eigenvalues ${E}_{jA}$ and
${E}_{kB}$)
of ${\cal H}_A$ and ${\cal H}_B$, respectively.
In this more general case the input is
not a maximally mixed state, but it is still a product state, e.g.\
in Eq.\,(23) the product of two separate
thermal states for $A$ and for $B$.

To recap: the key observation is that entangled states
produced by random processes have random relative phases, and when parts of
these states fall into a region with a nonlinear evolution map,
effectively the input to the map is
reduced to a product state, whose evolution is guaranteed to be
causal by the locality condition Eq.\,(16).
Put differently, a distant region
with a nonlinear evolution map might
cause random interference fringes throughout the observable
universe to shift; but random fringes
disappear when averaged over the varying relative
phases, therefore any shift in these fringes
also vanishes when averaged, leaving no observable effect of the map
in the external universe. In order to see any effect
of the nonlinear maps possibly lurking in distant extreme regions
of spacetime, it is necessary to design an experiment
which can send probe states out to infinity that are entangled
with laboratory states with known, coherently controlled phases.

If we denote the class of (local, completely positive)
linear maps by $\fL \subset \M
({\cal H})$, the complement (nonlinear maps) by $\NL= \M
\setminus \fL$, and the class of
signaling maps by $\fS$, we have just shown that $\fS \cap
\fL = \{ \}$.
%Another useful class of maps which we denote
%by $\F ({\cal H})$, and which our proposed experiment is sensitive to,
%are those ${\cal E}_{AB}$ for which the local components
%${\cal E}_A$ and ${\cal E}_B$ are linear
%(and completely positive). Any such map ${\cal E}_{AB}
%\in \F ({\cal H})$ can be written as
%\begin{equation}
%{\cal E}_{AB} ( \rho_{AB}) =
%{\cal E}_A \otimes {\cal E}_B (\rho_{AB}) + {\cal F} (\rho_{AB}) \; 
%\end{equation}
%in a unique way, where ${\cal F}$ is a smooth nonlinear map
%$W({\cal H}) \rightarrow W({\cal H})$ which satisfies ${\cal F} (\rho
%\otimes \sigma ) = 0$ for all
%$\rho \in S({\cal H}_A )$, $\sigma \in S({\cal H}_B )$,
%and ${\rm Tr}[{\cal F}(\rho_{AB})] = 0$ for all
%$\rho_{AB} \in S({\cal H})$. In~\cite{moretocome}, we will
%give a complete
%algebraic characterization of maps $\cal F$ satisfying
%these properties, and
%show that it is straightforward to produce both signaling
%and non-signaling examples of nonlinear maps in the
%$\F$-class~\cite{gisinetal}. Notice that the condition for a
%map ${\cal E}_{AB} \in \F ({\cal H})$ written in the form Eq.\,(21) to be
%signaling is simply that
%${\rm Tr}_B [{\cal F}(\rho_{AB})] \neq 0$ for some
%$\rho_{AB} \in S({\cal H})$.
In~\cite{moretocome}, we will give a complete
algebraic characterization of nonlinear maps satisfying
the locality condition Eq.\,(16), and
show that it is straightforward to produce both signaling
and non-signaling examples for maps in $\NL$~\cite{gisinetal};
that is, $\fS$ is a non-empty proper subset of $\NL$.
The evolution map $\cal E$ defined by Eqs.\,(11)--(12) is one example of
a class of local nonlinear maps---proposed by Horowitz and Maldacena
in~\cite{HM}
to describe quantum evolution through evaporating black holes---that
are signaling (i.e., belong to $\fS$). A detailed
analysis of the Horowitz-Maldacena
class of maps, along with a discussion of the motivation
for them, is what we turn to in the next Section.
%In general, the just-introduced
%subclasses $\fL ({\cal H})$, $\NL ({\cal H})$, and
%$\fS ({\cal H})$ of generalized
%quantum maps in $\M ({\cal H})$ obey the following set
%relationships (all
%inclusions are proper): $\NL = \M \setminus \fL \,$,
%$\fS \subset \NL \,$,
%$\F \cap \fS \neq \{ \} \,$, and $\F \cap (\NL \setminus \fS )
%\neq \{ \}$.

~~~~

\begin{center}
{\noindent
\bf 5. Nonlinear quantum maps associated with evaporating black
holes}
\end{center}

We begin with a brief review of the final-state boundary condition for
evaporating black holes as proposed in~\cite{HM} and further elucidated
in~\cite{GP}. In the semiclassical approximation,
the overall Hilbert space for the evaporation process can
be treated as a decomposition
\begin{equation}
{\cal H} = {\cal H}_M \otimes {\cal H}_F = {\cal H}_M
\otimes {\cal H}_{\rm in} \otimes {\cal H}_{\rm out} \; ,
\end{equation}
where ${\cal H}_M$ denotes the Hilbert space of the quantum field that
constitutes the collapsing body, and ${\cal H}_F$ is the Hilbert space
in which the quantum-field fluctuations around the background spacetime
determined by the ${\cal H}_M$ quantum state live. The separation of
${\cal H}$ into ${\cal H}_M$ and ${\cal H}_F$ reflects the semiclassical
nature of the treatment in a fundamental way. Moreover, the
fluctuation Hilbert space ${\cal H}_F$ can be further decomposed as
${\cal H}_F = \IN \otimes \OUT$, where $\IN$ and $\OUT$
denote the Hilbert spaces of
fluctuation modes confined inside and outside the event horizon,
respectively. Before evaporation, the quantum state $|\; \rangle
\in {\cal H}$ of the complete system  can be written as a product
\begin{equation}
|\; \rangle = |\psi_0 \rangle_M \otimes |0_U\rangle \; ,
\end{equation}
where $|\psi_0 \rangle_M \in {\cal H}_M$
is the initial wave function of the collapsing matter,
and the Unruh vacuum $|0_U \rangle \in 
{\cal H}_F = \IN \otimes \OUT$ is the
maximally entangled state
\begin{equation}
|0_U\rangle = \frac{1}{\sqrt{N}}
\sum_{k=1}^{N} | k_{\rm in} \rangle \otimes |k_{\rm out} \rangle
\; .
\end{equation}
Here $N$ is the common dimension (the number of degrees of
freedom necessary to completely describe the internal state of the black hole)
of all three Hilbert spaces ${\cal
H}_M$, $\IN$, and $\OUT$, and $\{ |k_{\rm in}\rangle \}$ and
$\{ |k_{\rm out} \rangle \}$, $k=1,2,\cdots,N$,
are fixed orthonormal bases for $\IN$ and $\OUT$, respectively.
After the hole evaporates completely, the ``final" Hilbert space is simply
$\OUT$, and, as we argued in Sec.\,1 above,
the usual semiclassical arguments inevitably imply a mixed
state $\rho_{\rm out}$ as the endpoint of complete evaporation (see
Fig.\,1 and the associated discussion), revealing
that the transition $|\psi_0 \rangle_M \mapsto \rho_{\rm out}$ is
manifestly non-unitary.

The Horowitz-Maldacena proposal (HM) imposes a boundary condition on the
final quantum state at the black-hole singularity by demanding that it
be equal to
\begin{equation}
|\Phi\rangle
\equiv U^{\dagger} \left[ \frac{1}{\sqrt{N}} \sum_{j=1}^{N} | j_M \rangle
\otimes |j_{\rm in} \rangle \right] \in {\cal H}_M \otimes \IN
\end{equation}
where $\{ |j_M \rangle \}$ is an orthonormal basis for ${\cal H}_M$,
and $U: {\cal H}_M \otimes \IN \rightarrow {\cal H}_M \otimes \IN$ is a
unitary transformation. More precisely, HM states the following:

\begin{quotation}
{\noindent \it There exists a unitary map
$U: {\cal H}_M \otimes \IN \rightarrow {\cal H}_M \otimes \IN$ such that
with $|\Phi\rangle \in {\cal H}$ defined as in Eq.\,(27),
the state $|\; \rangle$ [Eqs.\,(25)--(26)] evolves after
complete evaporation as
\begin{equation}
|\; \rangle \longmapsto \alpha \, P_{|\Phi\rangle \otimes \OUT}
\; | \; \rangle  \; ,
\end{equation}
where $\alpha \in \BR$ is a renormalization constant, and
$P_{|\Phi\rangle \otimes \OUT}$ denotes the projection onto the linear
subspace $|\Phi\rangle \otimes \OUT \equiv \{
|\Phi\rangle \otimes |v\rangle : \; |v\rangle \in \OUT \}$ of $\cal H$.}
\end{quotation}

The unitary operator $U$ describes the non-local evolution
of the black-hole quantum state near the singularity, as well as its
evolution in the semiclassical regime before the singularity; one would
expect a full quantum theory of gravity to be able
to completely specify this operator. To restore unitarity to the
transition map ${\cal H}_M \rightarrow \OUT$,
Horowitz and Maldacena~\cite{HM} further demand that $U$ be in
the form of a product corresponding to the absence
of entangling interactions between ${\cal H}_M$ and $\IN$:
\begin{equation}
U = S_1 \otimes S_2 \; ,
\end{equation}
where $S_1 : {\cal H}_M \rightarrow {\cal H}_M$ and $S_2 :
\IN \rightarrow \IN$ are unitary maps. To find the effective evolution
map ${\cal H}_M \rightarrow \OUT$ resulting from HM and the assumption
Eq.\,(29), start from the equality
\begin{equation}
\alpha \, P_{|\Phi\rangle \otimes \OUT} \;
(\, |\psi_0 \rangle_M \otimes |0_U\rangle \,)
= |\Phi\rangle \otimes |X_{\rm out} \rangle \; ,
\end{equation}
where $|X_{\rm out}\rangle$ is the state in $\OUT$ into which the initial
state $|\psi_0\rangle_M$ evolves after the evaporation. Contracting
both sides of Eq.\,(30) with $\langle \Phi |$ substituted from
Eq.\,(27)
\begin{eqnarray}
|X_{\rm out}\rangle & = & \frac{\alpha}{N}
\sum_{j=1}^{N}\langle j_M| \otimes \langle j_{\rm in}|
(S_1 \otimes S_2 ) 
|\psi_0\rangle_M \otimes \sum_{k=1}^{N} |k_{\rm in}\rangle \otimes
|k_{\rm out}\rangle \nonumber \\
& = & \frac{\alpha}{N} \sum_{j=1}^{N}
\sum_{k=1}^{N} \langle j_M|S_1 |\psi_0\rangle_M
\, \langle j_{\rm in} | S_2 | k_{\rm in} \rangle
\; |k_{\rm out} \rangle  \; .
\end{eqnarray}
In terms of the basis components
$X_{{\rm out} \; j} \equiv \langle j_{\rm out}|X_{\rm out}\rangle$,
$\psi_{0 \; k} \equiv \langle k_M | \psi_0 \rangle_M$,
$S_{1 \; jk} \equiv \langle j_M | S_1 | k_M \rangle$, and
$S_{2 \; jk} \equiv \langle j_{\rm in} | S_2 | k_{\rm in} \rangle$,
Eq.\,(31) can be rewritten in the matrix form
\begin{equation}
X_{{\rm out }\; k} = \frac{\alpha}{N}
\sum_{l=1}^{N} ({S_2}^{T} S_1 )_{\, kl} \; \psi_{0 \; l} \; \; ,
\end{equation}
where $S^{T}$ denotes matrix transpose of $S$. Since the transpose of a
unitary matrix is still unitary, Eq.\,(32) shows that (i) the
renormalization constant $\alpha =N$, and (ii) the transformation
$|\psi_0\rangle_M \mapsto |X_{\rm out}\rangle$ is unitary.

However, as pointed out by Gottesman and Preskill~\cite{GP}, entangling
interactions between ${\cal H}_M$ and $\IN$ are unavoidable in any
reasonably generic gravitational collapse scenario. Consequently,
we cannot expect the unitary operator $U$ to
have the product form Eq.\,(29) in general. For a general unitary map
$U: {\cal H}_M \otimes \IN
\rightarrow {\cal H}_M \otimes \IN$, the vector $|\Phi\rangle$
defined by Eq.\,(27) is an arbitrary element in
${\cal H}_M \otimes \IN$, and Eq.\,(31) leads to
the more general linear expression
\begin{equation}
X_{{\rm out }\; k} = {\alpha}
\sum_{l=1}^{N} T_{\, kl} \; \psi_{0 \; l} \; \; 
\end{equation}
instead of Eq.\,(32). Here $T$ denotes the matrix
\begin{equation}
T_{kl} \equiv \frac{1}{\sqrt{N}} \langle \, \Phi \; |l_M \rangle
\! \otimes \! | k_{\rm in} \rangle 
\end{equation}
which is unconstrained except for $\sum_{kl} |T_{kl}|^2 =
1/N$. Only when $U$ has the product form Eq.\,(29) $T$ equals ($1/N$
times) a unitary matrix [Eq.\,(32)].
Note that the constant $\alpha$ is to be determined from the
condition that $|X_{\rm out}\rangle$ remains normalized.
After this renormalization,
we can express the transformation ${\cal H}_M \rightarrow \OUT$
described by Eq.\,(33) more succinctly in the form
\begin{equation}
X_{{\rm out }\; k} = \frac{1}{(\, \sum_{i} | \! \sum_{j}
T_{\, ij} \; \psi_{0\; j} |^2 \, )^{\frac{1}{2}}}
\sum_{l=1}^{N} T_{\, kl} \; \psi_{0 \; l} \; \; 
\end{equation}
where now $T$ is a {\it completely unconstrained},
arbitrary matrix~\cite{ftnote3}.
While it maps pure states to pure states,
the transformation ${\cal H}_M \rightarrow \OUT$ specified by Eq.\,(35)
is not only nonunitary, but it is in fact {\it nonlinear};
linearity is recovered (along with unitarity) only when $T$ is
proportional to a unitary matrix.

In the preceding Sects.\,1--4, we argued that nonlinear quantum evolution
inside an evaporating black hole might have observable consequences outside
the event horizon when an entangled system (whose coherence is carefully
monitored) partially falls into the hole. In Sect.\,3
we proposed a specific
experiment that should be able to detect the presence of such
signaling nonlinear maps via terrestrial quantum interferometry. We
will now show that the HM-class of nonlinear maps defined in
Eq.\,(35) in fact belong to this signaling class quite generally.
Therefore, the HM boundary-condition proposal can
in principle be tested by terrestrial experiments.

Let us assume a causal configuration as depicted in Fig.\,1,
where a bipartite system $AB$ evolves to send its $B$-half
into a black-hole event horizon along a null geodesic, while the
$A$-half remains coherently monitored outside the horizon. We can
then further decompose the ``collapsing" Hilbert space ${\cal H}_M$ in
the form ${\cal H}_M = {\cal H}_A \otimes {\cal H}_B$, where ${\cal
H}_B$ now corresponds to all matter that falls into the black hole,
including the ``probe beam" $e$ of our trans-horizon Bell-correlation
experiment (cf.\ Fig.\,2
and the discussion following it in Sect.\,3), and ${\cal H}_A$
corresponds to all matter that remains outside the horizon, including
the interferometer beams which are monitored in the laboratory. We
also identify the outgoing Hilbert space $\OUT$ with ${\cal H}_M$, which
amounts to specifying a unitary map $U_M : {\cal H}_M \rightarrow \OUT$
connecting orthonormal basis sets in the two spaces. With this
identification, the ``evaporation" map ${\cal H}_M \rightarrow \OUT$ can
be treated simply as a map sending ${\cal H}_M$ onto ${\cal H}_M$.
Reinterpreted thus,
the action of a general quantum map in the class defined by Eq.\,(35)
can be written as
\begin{equation}
\rho_{AB} \longmapsto
\frac{T \, \rho_{AB} \, T^{\dagger}}
{{\rm Tr} (T \, \rho_{AB} \, T^{\dagger})} \; 
\end{equation}
on any state $\rho_{AB}$ in ${\cal H}_M
={\cal H}_A \otimes {\cal H}_B$,
where $T: {\cal H}_A \otimes {\cal H}_B
\rightarrow {\cal H}_A \otimes {\cal H}_B$ is a
(nonsingular) general linear
transformation~\cite{ftnote3}. To satisfy the locality condition as
formulated in Eq.\,(16), the map $T$ must have the
product form
\begin{equation}
T = T_A \otimes T_B \; ,
\end{equation}
where $T_A : {\cal H}_A \rightarrow {\cal H}_A$
and $T_B : {\cal H}_B \rightarrow {\cal H}_B$ are general linear maps.
Since subsystem $A$ remains outside the event horizon, the evolution map
$T_A$ must remain unitary, and we can assume
(for simplicity and without loss of
generality) that $T_A = \mathbb{I}_A$. Then the quantum evolution
map Eq.\,(36) acting
on the Hilbert space ${\cal H}_M = {\cal H}_A \otimes {\cal H}_B$
takes the more transparent form
\begin{equation}
{\cal E}_{AB} \; : \; \rho_{AB} \longmapsto
\frac{{\pmb 1}_A \otimes {\cal T}_B \; ( \rho_{AB})}
{{\rm Tr} [{\pmb 1}_A \otimes {\cal T}_B \; ( \rho_{AB})]} \; ,
\end{equation}
where ${\pmb 1}_A = {\cal E}_A$ denotes the identity
map on states of ${\cal H}_A$, and
${\cal T}_B$ denotes the linear transformation (not a quantum map)
\begin{equation}
{\cal T}_B \; : \; \rho_{B} \longmapsto
T_B \, \rho_{B} \, {T_B}^{\dagger} \; 
\end{equation}
on states of ${\cal H}_B$. When $\rho_{AB}$ is a product state
$\rho_{AB}=\rho_A \otimes \rho_B$, the action of
${\cal E}_{AB}$ has the manifestly local form
\begin{equation}
{\cal E}_{AB}(\rho_{AB})= {\cal E}_A (\rho_A )
\otimes {\cal E}_B (\rho_B ) \; ,
\end{equation}
where ${\cal E}_A = {\pmb 1}_A$, and ${\cal E}_B$ is the nonlinear
quantum map
\begin{equation}
{\cal E}_{B} \; : \; \rho_{B} \longmapsto
\frac{{\cal T}_B (\rho_B)}{{\rm Tr}_B [{\cal T}_B (\rho_B )]}
=
\frac{T_B   \rho_{B} {T_B}^{\dagger}}
{{\rm Tr}_B ( T_B   \rho_{B} {T_B}^{\dagger})} \; 
\end{equation}
mapping ${\cal H}_B$-states onto ${\cal H}_B$-states (compare
Eq.\,(40) with Eq.\,(16) above in Sect.\,4). By contrast,
when $\rho_{AB}$ is
entangled the action of ${\cal E}_{AB}$ does not
have the simple product form of Eq.\,(40).

In Sect.\,4 above, we identified the
criterion for a quantum map ${\cal E}_{AB}$ to be signaling
[see Eq.\,(17) and the associated discussion] as simply the condition that
\begin{equation}
{\rm Tr}_B \, [ \, {\cal E}_{AB} (\rho_{AB}) \, ]
\neq {\cal E}_A \, [ \, {\rm Tr}_B ( \rho_{AB}  ) \, ] \; 
\end{equation}
for {\it some} (necessarily entangled) state $\rho_{AB}$. Now consider
a class of entangled states
$\rho_{AB}$ in the form of a convex linear combination
\begin{equation}
\rho_{AB} = \lambda_1 \, \rho_A \otimes \rho_B + \lambda_2
\, \sigma_A \otimes \sigma_B \; ,
\end{equation}
where $\rho_A , \; \sigma_A$ , and $\rho_B , \; \sigma_B$ are
(normalized) states in
${\cal H}_A$ and ${\cal H}_B$, respectively, and $\lambda_1 >0,
\; \lambda_2 > 0 , \; \lambda_1 + \lambda_2 = 1$ are real coefficients.
Introduce the real numbers
\begin{eqnarray}
n_1 & \equiv & {\rm Tr}_B [{\cal T}_B (\rho_B )] = {\rm Tr}_B (T_B \rho_B
{T_B}^{\dagger}) \; , \nonumber \\
n_2 & \equiv & {\rm Tr}_B [{\cal T}_B (\sigma_B )] = {\rm Tr}_B (T_B \sigma_B
{T_B}^{\dagger}) \; .
\end{eqnarray}
The right-hand-side of Eq.\,(42) is simply ${\rm Tr}_B (\rho_{AB})$
(recall that ${\cal E}_A = {\pmb 1}_A$):
\begin{equation}
{\cal E}_A \, [ \, {\rm Tr}_B ( \rho_{AB}  )]
= \lambda_1 \rho_A + \lambda_2 \sigma_A \; ,
\end{equation}
while the left-hand-side is
\begin{equation}
{\rm Tr}_B \, [ \, {\cal E}_{AB} (\rho_{AB}) \, ]
= \frac{\lambda_1 \, n_1 \, \rho_A \; + \; \lambda_2 \, n_2 \, \sigma_A}
{\lambda_1 \, n_1 + \lambda_2 \, n_2} \; .
\end{equation}
But
\begin{equation}
\lambda_1 \rho_A + \lambda_2 \sigma_A \; \neq \;
\frac{\lambda_1 \, n_1 \, \rho_A \; + \; \lambda_2 \, n_2 \, \sigma_A}
{\lambda_1 \, n_1 + \lambda_2 \, n_2}
\end{equation}
{\it unless} at least one of the conditions: (i) $n_1 = n_2$, or
(ii) $\rho_A=\sigma_A$ holds. The condition (i) does not hold in general
unless the linear operator $T_B$ is unitary (or a scalar multiple of a
unitary operator), and condition (ii) does not hold in general unless
$\rho_{AB}$ is a product state. Therefore the nonlinear quantum map ${\cal
E}_{AB}$ defined by Eqs.\,(38)--(39) is in general in the signaling
class. A {\em specific} example of a map in this class, and the signal
that it produces in the Zou-Wang-Mandel interferometer of our proposed
experiment, were described in Sect.\,3 above [cf.\
Eqs.\,(11)--(13)].

~~~~~

\begin{center}
{\noindent
\bf 6. Experimental prospects for the detection
of distant nonlinearities via terrestrial probes}
\end{center}

We have shown that the quantum
maps which likely characterize quantum evolution through
evaporating black holes according to the Horowitz-Maldacena~\cite{HM}
boundary-condition proposal are a signaling class.
It is clear that the HM-class of maps are
in principle detectable with the apparatus
of our proposed Bell-correlation
experiment, namely the Zou-Wang-Mandel (ZWM)
interferometer depicted in Fig.\,2
[for the specific example of Eqs.\,(11)--(13),
the detection signal is a $45^{\circ}$ shift
in the detector's interference fringes].
What are the practical prospects for actually detecting
the presence of signaling nonlinear maps inside
black-holes, assuming such maps do indeed exist?
The set of phenomena which may impede detection
efficiency in the trans-horizon Bell-correlation experiment can
be naturally divided into two types according to whether they take place
inside or outside the event horizon.

Inside the event horizon, the precise nature of the signal
produced in the ZWM interferometer when the probe beam is sent
into an evaporating hole will depend on the nature of the unitary
operator $U$ characterizing the HM boundary condition, Eq.\,(27).
If, as predicted~\cite{HM,GP},
the operator $U$ involves nonlocal phases which oscillate chaotically
at Planckian frequencies near the singularity, then each ZWM
photon entering the hole is likely to experience a different nonlinear
evolution map ${\cal E}_{AB}$, and the observed signal will be an
average over such maps. Preliminary calculations show
that this averaging will affect the local interference pattern back in
the laboratory by erasing relative phases and thus drastically reducing
fringe visibility. The ultimate result of the fluctuations
is a complete erasure of the interference pattern [Eqs.\,(6)].
While this erasure gives rise
to a qualitatively different signal than Eq.\,(13), the absence
or reduction of interference still constitutes a strong detection signal
in the ZWM setup; in other words, signal strength, i.e.\
the capability of the ZWM instrument to terrestrially detect
nonlinear maps, is not diminished. In the ZWM interferometer,
``no detection" corresponds to the {\it presence}
of the specific interference pattern Eqs.\,(6); any deviation from this
two-dimensional data set (i.e.\ the quantity $p_A - p_B$ as a function
of the angles $\phi$ and $\theta = \arccos |a|$) would register
as a detection of nonlinear quantum evolution along the probe beam,
since no other physical phenomena can give rise to such deviations.
This robustness of the detection signal
in the presence of a rapidly fluctuating nonlinear map
is a distinguishing feature of the ZWM instrument, and places it in
contrast with other possible experimental designs to search for
nonlinear quantum evolution at large distances.
Clearly, the ZWM interferometer is not the only
instrument capable of probing distant nonlinearities in quantum
mechanics; any entangled system can in principle be used as
such a probe. However, the second-order interference
[i.e.\ the specific interference pattern Eqs.\,(6)
between the $u$--$d$ beams which is maintained after
tracing over the probe beam's Hilbert space ${\cal H}_e$]
makes the ZWM interferometer more sensitive to violations of linearity
along the probe beam than other, more obvious experiments one can
design. For example, when relative phases are carefully
controlled, an EPR pair
of polarization-entangled photons can be used as a
simple detector by monitoring the polarization
statistics of the local, ``laboratory" photon while the other, ``probe"
photon escapes to infinity. Indeed, it can be shown~\cite{moretocome}
that precisely because of the expected fluctuating geometry inside
the event horizon, such a detector {\em cannot} detect HM-maps of
the type Eqs.\,(38)--(39), in contrast with the ZWM detector which can.
A detailed quantitative
analysis of the interferometric response to fluctuating nonlinear
maps will be given in the forthcoming paper~\cite{moretocome}.

Outside the event horizon, environmental decoherence of the probe beam
at large distances places other fundamental limits
on the visibility of any deviations from the standard
``unitary" interference signal Eqs.\,(6)
in our proposed experiment.
Additional limits arise from the diffraction of the probe beam at large
distances, which will let only a small fraction of the beam's
``footprint" to impact a black hole. There are also obvious difficulties
with targeting specific black-hole candidates which might limit
the ultimate experiment to a global search across the sky.
A detailed analysis of these limits will be given in~\cite{moretocome}.
At the very least, an easier to obtain, but perhaps less interesting,
result of the experiment would be to place novel limits on possible
nonlinearities~\cite{nlrefs,weinberg} in the quantum
evolution of the probe beam $e$ as it propagates through free space.

\begin{center}
{\noindent
\bf Acknowledgements}
\end{center}

The research described in this paper was carried out
at the Jet Propulsion Laboratory under a contract with the National
Aeronautics and Space Administration (NASA), and
was supported by grants from NASA and the Defense
Advanced Research Projects Agency. We would like to thank our colleagues
in the Quantum Computing Technologies group at JPL, in particular
Jonathan Dowling and Chris Adami, for stimulating discussions.

~~~~


\begin{thebibliography}{5}
\bibitem{ftnote2} For maps sending pure states to pure
 states, linearity (plus probability conservation) is equivalent
 to unitarity.
\bibitem{ftnote3} The only constraint on the linear operator $T$
 entering the construction Eq.\,(35) or Eq.\,(36) is that
 $T$ must not have zero as an eigenvalue (e.g., projection-like
 operators are excluded), since with a singular
 $T$ this construction fails to give a well-defined quantum map.
\end{thebibliography}
\end{document}